**Yield Strength as a Thermodynamic Consequence of Information Erasure**

Parag Katira[1] and Henry Hess[1*]

[1] Department of Biomedical Engineering, Columbia University, USA

Corresponding Author: Email – hhess@columbia.edu

**We observe that the yield strength of a variety of materials, including highly structured and densely packed metals, alloys and semi-crystalline polymers is reasonably approximated by the thermal energy density of the material. This suggests that it is related to the entropic cost of the irreversible work done during plastic deformation rather than the enthalpic cost that depends on the elastic modulus of the material. Here we propose that the entropic cost of material rearrangement in crystalline solids arises from the difference in the uncertainty in building block positions before and after yielding and estimate it using Landauer's principle for information processing. The yield strength thus obtained is given by the thermal energy density of the material multiplied by ln(2) and provides a guidepost in estimating the strength of materials complementary to the "theoretical strength of solids".**

Plastic deformation, marked by an irreversible change in the shape of a ductile material, occurs under external stresses exceeding the yield strength of the material. Plastic deformation is a mode of mechanical failure in materials and consequently the yield strength of materials is a key property considered during materials design and selection. The origins of plasticity lie in the movement of atoms or molecules past one another in a material under stress, and attempts to calculate the yield strength are based on predicting the amount of energy required for these rearrangements.[1,2] The theoretical strength of materials corresponds to the energy required to overcome the attractive forces between the atomic or molecular building blocks (the smallest independently movable element) of the material and is the only widely used benchmark to compare experimental yield strength values against.[2-4] For a perfectly crystalline material, this estimate is approximately given by $E/\pi$, where E is the Young's modulus of the material.[2,5] However, the theoretical strength is typically two to three orders of magnitude larger than the experimentally obtained yield strength for most materials used in engineering. The presence of defects within the crystal lattice such as dislocations, twin boundaries and grain boundaries can greatly reduce the energy required to overcome the attractive forces between building blocks, and a significant portion of plasticity theory is devoted to studying the effect of these defects on a material's yield strength.[6] This has resulted in many different models that qualitatively explain how various processing methods, which change for example grain size or defect density, affect the yield strength of materials.[7] However, due to the many mechanisms for producing and moving defects, plasticity theory is not able to quantitatively predict the yield strength of a material based solely on its composition,[6] but rather relies on empirically determined parameters.[4,8]

Here we derive a second guidepost complementing the theoretical strength by focusing on the minimal entropic cost of irreversible rearrangement of building blocks of a solid under a uniaxial stress. The entropic contribution arises from the difference in the uncertainty of building block positions (given by the Shannon entropy[9]) before and after the application of stress. An increase in the Shannon entropy of a system requires the dissipation of a minimum amount of energy according to the second law of thermodynamics and thereby provides a lower limit for the applied stress required to permit plastic deformation. This is an application of Landauer's principle originally formulated with respect to computations, stating that any logically irreversible manipulation of information, such as the erasure of a bit, must be accompanied by a corresponding entropy increase in non-information bearing degrees of freedom of the information processing apparatus or its environment, i.e. energy dissipation into the surroundings as heat.[10] In this case, we regard the positions of the building block of a material to constitute information, and an increase in the uncertainty in building block positions as information erasure. The application of information theory and Landauer's principle simplifies an extremely complicated problem of tracking the phase space available to every particle in the system as it is plastically deformed and helps us obtain a first order estimate of the entropic cost of plastic deformation. It is complementary to the theoretical strength of the material which focuses on the enthalpic cost of material rearrangement.

To derive the "entropic yield strength", consider a solid specimen being deformed along one of its axis by the application of a tensile force[11] inside a temperature bath (the specimen and the bath constitute the "system" here). The knowledge of the original shape of the solid specimen implies the knowledge of the positions for all of its building blocks. Under tension, the solid specimen deforms and the building blocks shift positions accordingly. If the specimen deforms only elastically, upon releasing the tensile force, the building blocks will return to their original positions and no information is gained or lost. However, if the specimen has deformed plastically, some of its buildings blocks have moved to new positions. One of the main assumptions of plastic deformation is that there is no net change in the volume of the solid specimen. Hence, every building block that has receded from the original surface of the specimen in a direction orthogonal to the applied stress is matched by a building block crossing over the original surface at either end of the specimen along the direction of the applied stress (Figure 1). If $N$ building blocks are thus rearranged the increase in the positional uncertainity of the building blocks, quantified by the change in the Shannon entropy of the system is

$$\Delta H = \ln(2^N), \qquad (1)$$

because the $N$ building blocks can be evenly divided in $2^N$ ways between the two ends of the specimen. This is equivalent to a loss of $N$ bits of information[12] regarding the positions of the building blocks. Landauer's principle requires that this loss of information must result in a dissipation of heat in an amount of at least $k_bT\Delta H$ into the bath.[13,14] This dissipated heat is equal to the amount of irreversible work done during the process. This work is given by $\sigma\Delta V$, where $\sigma$ is the stress applied to the solid and $\Delta V=NV_b$ is the total volume change along the direction of the applied stress with $V_b$ being the volume of a single building block. From the inequality $\sigma\Delta V \geq k_bT\Delta H$ and equation (1), we obtain

$$\sigma \geq \frac{k_BT\ln(2)}{V_b} \qquad (2)$$

By replacing $V_b$ with $M_w/\rho$, where $M_w$ is the molecular weight of each of the building blocks and $\rho$ is the density of the material, we obtain

$$\sigma \geq \frac{\rho k_BT\ln(2)}{M_w} \qquad (3)$$

This equation gives the minimum amount of stress required to plastically deform the specimen based on the entropic cost of material rearrangement and is thus the "entropic yield strength" of the material.

The result (Eq. 3) obtained from the analysis of the entropy change of the system can also be obtained by looking at the rearrangement of individual building blocks during plastic deformation (Figure 2). When a specimen is strained, its internal energy increases on account of the increased potential energy of the building blocks interacting with each other via attractive bonds. The internal energy of the system at a particular strain can be lowered by the addition of a building block along the direction of the strain, which decreases the amount of stretch per bond. This can occur by a building block at the surface of the material moving into the bulk of the material from a direction orthogonal to the strain in the material. However, the surface block entering the bulk needs to insert either to the left or the right of the building block under it. This choice made by the displaced building block shrinks the available phase space and results in a decrease in the conformational entropy of the displaced building block similar to that observed during symmetry breaking[15] or decision making by a molecular machine.[16] The displacement of the building block either to the left or the right changes the entropy of the building block by[15] $k_B\ln(p_i)$, where $p_i$ is the probability of it being confined in state $i$.[15] In this case $p_i = p_{left} = p_{right} = 0.5$ for an isotropic system. Thus, the rearrangement of a surface building block into the bulk results in an entropy production of $-k_B\ln(2)$. For this rearrangement to take place spontaneously, according to the second law of thermodynamics the change in the free energy $\Delta F$ has to be negative, i.e. $\Delta E - T\Delta S \leq 0$. Thus, the energy dissipated by the system during each rearrangement of the building blocks is given by $\Delta E \leq -k_BT\ln(2)$ and is extracted from the work done on the system. Since this operation increases the volume of the material by that of one building block along the direction of the stretch, the work done is given by $\sigma V_b$ where $V_b$ is the volume of a building block as defined earlier. From this and using again $V_b=M_w/\rho$, we arrive at the same expression for the amount of stress required to cause yielding (Eq. 3).

The close agreement of experimentally observed yield strength values for a wide range of materials considered in the Ashby Plot of yield strength vs. material density of Figure 3 suggests the universality of this argument in predicting the yield strength of ductile materials. Here, we assumed the building blocks to be the predominant atom in metal alloys or the mers in the polymers. The ceramic materials, composites and foams shown in the original Ashby Plot are not included in Figure 3 since they do not typically exhibit plasticity. It also suggests that while the enthalpic cost of building block rearrangement may influence the kinetics of yielding, the entropic cost of rearrangements has a major contribution to the yield strength of a typical engineering material.

The experimental observation of lower yield strengths than those predicted by Equation (4) can originate from collective movements, where the "building block" is effectively larger than the atomic constituents. Factors that influence the initial and final uncertainties in building block positions in a material will also influence the extent of the entropic contribution to yield strength. For a perfectly isotropic, defect free material, the entropic yield strength is given by $k_BT\ln(2)/V_b$. However, the presence of specific defects that increase the likelihood of movement of the building blocks in one particular direction will reduce the uncertainty in final positions and thus lower the entropic contribution to the yield strength. Also, factors such as temperature, which can change the initial uncertainty in building block positions, may influence the entropic contribution to yield strength.

It has been previously pointed out that the yield strength of hard-sphere glasses,[17] emulsions[18,19] and certain metallic glasses[20] is related to the ratio $k_bT/V_b$. For hard sphere glasses and other glassy materials, $k_bT$ corresponds to the entropic barrier from thermal collisions between neighbors and the available free volume into which a building block can be moved to is on the order of $V_b$. However, the entropic yield strength of $k_BT\ln(2)/V_b$ discussed here is derived from Landauer's principle rather than a consideration of steric barriers to the movement of building blocks.

Yield strength prediction based on the entropy of building block rearrangements, as described here, provides a lower bound on the yield strength of materials complementary to the theoretical strength of materials, which is the upper bound on the yield strength of materials. For a rearrangement where no new surfaces are created or the average number of internal bonds remains the same, the entropic costs of rearrangements exceeds the negligible enthalpic cost of rearrangement. While the formulation of the model used to arrive at equation (3) is based on a number of assumptions that simplify an extremely complex system, it provides a new, general paradigm to examine the yield strength of materials.

## References

1   Orowan, E. Problems of plastic gliding. *Proceedings (Physical Society (Great Britain) : 1926)* **52**, 8-22, doi:10.1088/0959-5309/52/1/303 (1940).

2   Bragg, L. A theory of the strength of metals. *Nature* **149**, 511-513, doi:Doi 10.1038/149511a0 (1942).

3   Macmillan, N. H. Review: The Theoretical Strength of Solids. *Journal of Materials Science* **7**, 239-254 (1972).

4   Cottrell, A. H. Theory of dislocations. *Progress in Metal Physics* **1**, 77-126, doi:http://dx.doi.org/10.1016/0502-8205(49)90004-0 (1949).

5   Hofmann, D. C. *Designing Bulk Metallic Glass Matrix Composites with High Toughness and Tensile Ductility* Ph.D. thesis, California Institute of Technology, (2009).

6   Lubliner, J. *Plasticity theory*.  (Courier Dover Publications, 2008).

7   Read, W. T. *Dislocations in crystals*.  (McGraw-Hill, 1953).

8   Wang, J. S., Mulholland, M. D., Olson, G. B. & Seidman, D. N. Prediction of the yield strength of a secondary-hardening steel. *Acta Mater* **61**, 4939-4952, doi:DOI 10.1016/j.actamat.2013.04.052 (2013).

9   Shannon, C. E. A Mathematical Theory of Communication. *At&T Tech J* **27**, 623-656 (1948).

10  Landauer, R. Irreversibility and Heat Generation in the Computing Process. *Ibm J Res Dev* **5**, 183-191 (1961).

11   *Tension Test*, <http://www.instron.us/wa/glossary/Tension-Test.aspx> (2007).

12   Schneider, T. D. Theory of molecular machines. I. Channel capacity of molecular machines. *J Theor Biol* **148**, 83-123, doi:Doi 10.1016/S0022-5193(05)80466-7 (1991).

13   Leff, H. S. & Rex, A. F. *Maxwell's demon : entropy, information, computing*. (Princeton University Press, 1990).

14   Bennett, C. H. Notes on Landauer's principle, reversible computation, and Maxwell's Demon. *Stud Hist Philos M P* **34B**, 501-510, doi:Doi 10.1016/S1355-2198(03)00039-X (2003).

15   Roldan, E., Martinez, I. A., Parrondo, J. M. R. & Petrov, D. Universal features in the energetics of symmetry breaking. *Nat Phys* **10**, 457-461, doi:Doi 10.1038/Nphys2940 (2014).

16   Schneider, T. D. Theory of molecular machines. II. Energy dissipation from molecular machines. *J Theor Biol* **148**, 125-137, doi:Doi 10.1016/S0022-5193(05)80467-9 (1991).

17   van der Vaart, K. *et al.* Rheology of concentrated soft and hard-sphere suspensions. *J Rheol* **57**, 1195-1209, doi:Doi 10.1122/1.4808054 (2013).

18   Schall, P., Weitz, D. A. & Spaepen, F. Structural rearrangements that govern flow in colloidal glasses. *Science* **318**, 1895-1899, doi:DOI 10.1126/science.1149308 (2007).

19   Mason, T. G., Bibette, J. & Weitz, D. A. Yielding and flow of monodisperse emulsions. *J Colloid Interf Sci* **179**, 439-448, doi:DOI 10.1006/jcis.1996.0235 (1996).

20   Wang, G. *et al.* Correlation between elastic structural behavior and yield strength of metallic glasses. *Acta Mater* **60**, 3074-3083, doi:DOI 10.1016/j.actamat.2012.02.012 (2012).

21   Ashby, M. *Materials Selection in Mechanical Design*. 3 edn, (Pergamon Press, 1992).

**Acknowledgments**

H.H. thanks Thomas Schneider for the introduction to information theory, Veronica Reynolds for her assistance in researching materials properties, and Ofer Idan, Takahiro Nitta, and Sarah Marzen for helpful discussions. This material is based upon work supported by, or in part by, the U.S. Army Research Laboratory and the U.S. Army Research Office under contract number W911NF-13-1-0390.

**Conflicts of Interest**

The authors have no competing financial interests.

**Figures**

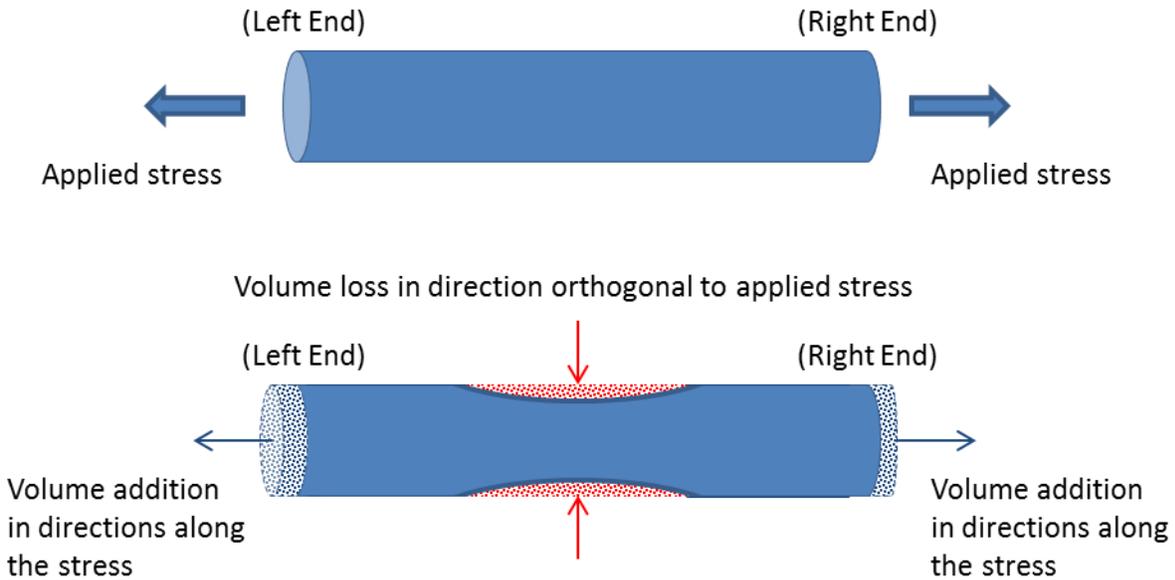

**Figure 1: Material plastically deformed by a tensile stress.** Rearrangement of N building blocks from a direction orthogonal to the applied stress to either end of the material along the applied stress occur in $2^N$ ways that result in the same NV volumetric strain along the direction of the applied stress. The irreversible work done in this process is then given by $\sigma$ NV, and the Shannon Entropy generated during this process is $\ln(2^N)$ (or N bits).

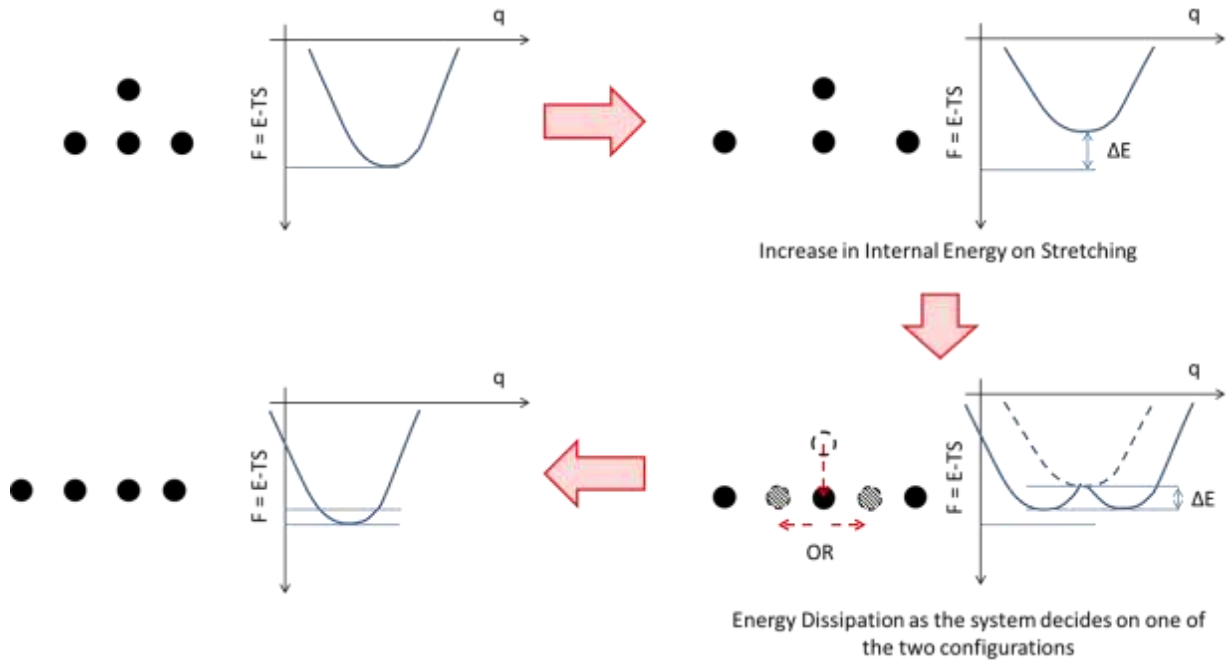

**Figure 2**. **Energetics of building block rearrangements at the microscale**. The instantaneous confinement of the system in one of the two possible lower energy states has characteristics similar to a symmetry breaking operation and is associated with an energy dissipation of at least $k_B T \ln(p_i)$ where $p_i$ is the probability of the system ending up in state i.

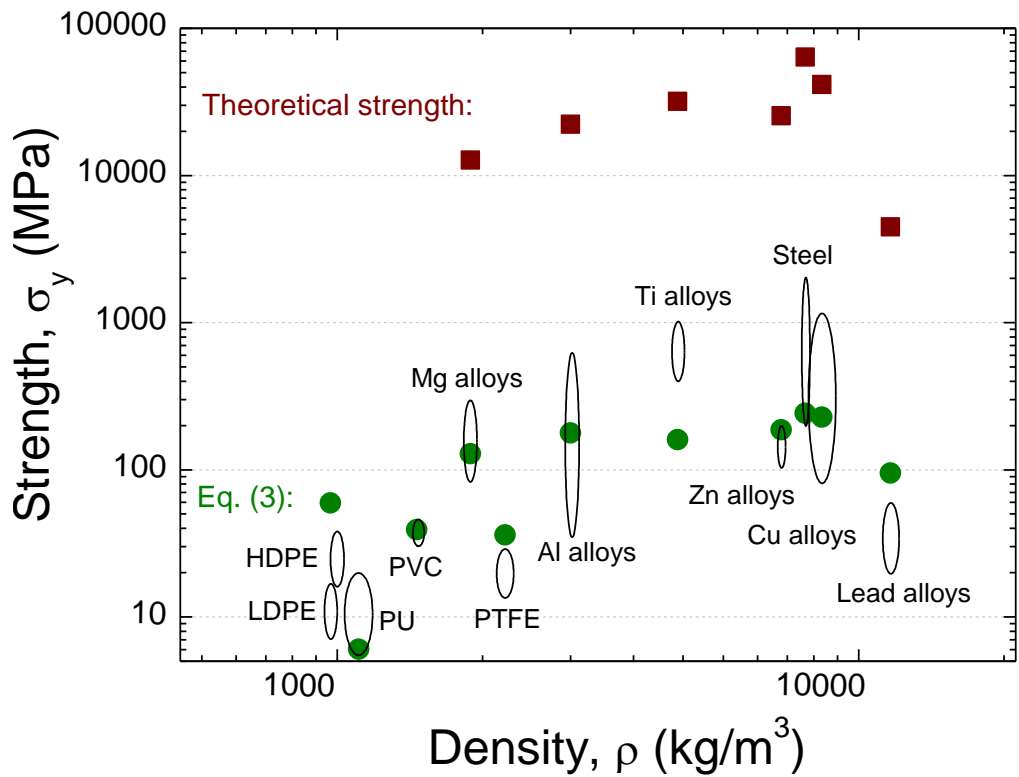

**Figure 3: Ashby Plot of yield strength as function of density for a variety of materials.** The range of yield strengths $\sigma_y$ of typical materials (adapted from M. Ashby: "Materials Selection in Mechanical Design", 1992, Fig. 3.18),[21] the entropic yield strength calculated from Eq. (3) (green circles), and the theoretical strength estimated by $E/\pi$ (red squares). Materials parameters (density $\rho$, molecular weight $M_W$, Young's modulus E): High-density/low-density Polyethylene (HDPE/LDPE) – 970 kg/m$^3$, 28 Da, 0.8 GPa; Polyurethane (PU) - 1100 kg/m$^3$, 312 Da, 1.6 GPa; Polyvinylchloride (PVC) - 1420 kg/m$^3$, 62 Da, 3 GPa; Magnesium alloys - 1800 kg/m$^3$, 24 Da, 40 GPa; Polytetrafluoroethylene (PTFE) - 2100 kg/m$^3$, 100 Da, 0.5 GPa; Aluminum alloys - 2800 kg/m$^3$, 27 Da, 70 GPa; Titanium alloys – 4500 kg/m$^3$, 48 Da, 100 GPa; Zinc alloys - 7100 kg/m$^3$, 65 Da, 80 GPa; Steel – 7900 kg/m$^3$, 56 Da, 200 GPa; Copper alloys – 8500 kg/m$^3$, 64 Da, 130 GPa; Lead alloys – 11500 kg/m$^3$, 207 Da, 14 GPa.